\documentclass[a4paper,11pt]{article}
\usepackage{pos}

\usepackage{graphicx}
\usepackage{subcaption}
\usepackage{cleveref}
\usepackage{braket}
\usepackage{slashed,bbm}

\title{QED with massive photons for precision physics: zero modes and first result
for the hadron spectrum}
\ShortTitle{QED$_{\textrm{M}}$}

\author[a]{M. A. Clark}
\author[b]{M. Della Morte}
\author[c]{Z. Hall}
\author[d]{B. H{\"o}rz}
\author[c,d]{A. Nicholson}
\author*[e,f]{A. Shindler}
\author*[b]{J.~T.~Tsang}
\author[d,g]{A.~Walker-Loud}
\author[h,g,i]{H. Yan}

\affiliation[a]{NVIDIA Corporation, Santa Clara, CA 95050, USA}
\affiliation[b]{IMADA \& CP$^3$-Origins. University of Southern Denmark.
  Campusvej 55, DK-5230 Odense, Denmark}
\affiliation[c]{Department of Physics and Astronomy, University of North Carolina, Chapel Hill, NC 27516-3255, USA}
\affiliation[d]{Nuclear Science Division, Lawrence Berkeley National Laboratory, Berkeley, CA 94720, USA}
\affiliation[e]{Department of Physics and Astronomy,
    Michigan State University,
    East Lansing, 48824, Michigan, USA}
\affiliation[f]{Facility for Rare Isotope Beams,
    Michigan State University,
    East Lansing, 48824, Michigan, USA}
\affiliation[g]{Department of Physics,
    University of California,
    Berkeley, CA 94720, USA}
\affiliation[h]{School of Physics,
    Peking University, Beijing 100871, China}
\affiliation[i]{School of Physics,
    Xi'an Jiaotong University, Xi'an 710049, China}

\emailAdd{shindler@frib.msu.edu}
\emailAdd{tsang@imada.sdu.dk}

\abstract{ The current precision reached by lattice QCD calculations of
  low-energy hadronic observables, requires not only the introduction of
  electromagnetic corrections, but also control over all the potential
  systematic uncertainties introduced by the lattice version of QED.
  Introducing a massive photon as an infrared regulator in lattice QED, provides
  a well defined theory, dubbed QED$_{\textrm{M}}$, amenable to numerical
  evaluation~\cite{Endres:2015gda}.  The photon mass is removed through
  extrapolation. In this contribution we scrutinise aspects of
  QED$_{\textrm{M}}$ such as the presence and fate of the zero modes
  contributions and we describe the determination of the photon mass corrections
  in finite and infinite volume. We demonstrate that the required extrapolations
  are well controlled using numerical data obtained on two ensembles which only
  differ in volume.}

\FullConference{
The 38th International Symposium on Lattice Field Theory, LATTICE2021
26th-30th July, 2021
Zoom/Gather@Massachusetts Institute of Technology
}

\begin{document}
\maketitle

\section{Introduction}
\label{sec:intro}
Lattice QCD (LQCD) calculations are reaching, for several observables, a control
over statistical and systematic effects of the order of the percent and below.
An incomplete list includes the recent calculations of radiative leptonic decays
of light mesons~\cite{Giusti:2017dwk,Frezzotti:2020bfa} the baryon
spectrum~\cite{Borsanyi:2014jba}, $g_A$~\cite{Chang:2018uxx} and
$(g-2)_\mu$~\cite{Borsanyi:2020mff}~\footnote{This calculation includes QED
  corrections using the so-called QED$_{\textrm{L}}$~formulation.}. Isospin-breaking (IB)
corrections include the difference between the up and down quark masses, $\delta
m = m_u - m_d$, and the electromagnetic interactions between quarks proportional
to the electromagnetic coupling $\alpha_{\rm{em}}$. Both IB sources provide,
usually, corrections of the order of the percent, as O($\delta m
/\Lambda_{\rm{QCD}}$) or O($\alpha_{\rm{em}}$). A sub-percent uncertainty in any
LQCD calculation can only be reached if we control IB corrections.

In this contribution we discuss QED corrections. Discretising
QED on a finite volume lattice $V = L^3 \times T$, of sides $L_\mu=(L,L,L,T)$,
hides subtleties that need to be taken into
account for a robust determination of QED corrections to LQCD calculations.
QED is usually discretised in a finite volume, see Ref.~\cite{Blum:2017cer} for a different approach,
and the standard choice is to use periodic boundary conditions for the gauge fields.
A notable exception is the use of $C^*$ boundary conditions proposed in Ref.~\cite{Lucini:2015hfa},
a formulation called QED$_{\textrm{C}}$. One of the obstacles that prevent a straightforward
implementation of QED on the lattice is the propagation of charged states.
It is easy to convince oneself that with periodic boundary conditions
charged states in the Hilbert space of the theory are forbidden by Gauss'
law~\cite{Creutz:1978xw,Hayakawa:2008an}.
Given the fact that charged particles are not gauge invariant, one might
presume that a local gauge fixing condition would solve the problem, but that is not true.
The reason is that, in a finite volume, even after a local gauge fixing procedure,
there still exists a gauge transformation, not connected with the identity,
that is still a symmetry of the theory: the so-called large gauge transformations (LGT).
If we denote the photon and fermion fields respectively with $A_\mu(x)$ and $\psi(x),~\overline{\psi}(x)$
the LGT read
\begin{eqnarray}
A_\mu(x) &\rightarrow& A_\mu'(x) = A_\mu(x) + 2 \pi \frac{r_\mu}{L_\mu} \nonumber \\
\psi(x) &\rightarrow& \psi'(x) = \rm{e}^{i 2\pi r_\mu \left(\frac{x}{L}\right)_\mu} \psi(x) \nonumber \\
\overline{\psi}(x) &\rightarrow& \overline{\psi}'(x) =  \overline{\psi}(x)\rm{e}^{-i 2\pi r_\mu \left(\frac{x}{L}\right)_\mu}\,,
\end{eqnarray}
where $r_\mu \in {\mathbbm{Z}}^4$ and $(x/L)_\mu = (x_1/L,x_2/L,x_3/L,x_4/T)$.
LGT are a consequence of the thermal boundary conditions imposed on the fermion fields, and
any QED formulation on the lattice has to provide a solution for the breaking of LGT.
The solution of this problem relates to the analysis of the zero modes of the photon field.

Most of the solutions proposed, like QED$_{\textrm{TL}}$~\cite{Duncan:1996xy} or QED$_{\textrm{L}}$~\cite{Hayakawa:2008an}, remove the zero modes "by hand"
introducing potential non-localities in the theory. A discussion on the fate of the
zero modes for the most popular QED formulations on the lattice can be found in Ref.~\cite{Patella:2017fgk}.

In Ref.~\cite{Endres:2015gda} we have proposed a discretisation of lattice QED, that introduces
a non-zero photon mass, $m_\gamma \neq 0$, and labelled as QED$_{\textrm{M}}$.
A first exploratory study of the isospin breaking corrections to $g-2$ computed with QED$_{\textrm{M}}$ and QED$_{\textrm{L}}$ has been presented in~\cite{Bussone:2017xkb}.
In the next section we review some of the properties of QED$_{\textrm{M}}$~
in relation with the fate of the zero modes of the photon.

\section{QED$_{\textrm{M}}$}
\label{sec:qedm}

In Ref.~\cite{Endres:2015gda} we have introduced a new lattice QED action (QED$_{\textrm{M}}$)
\begin{equation}
S_{\rm{M}} =  a^4 \sum_x \frac{1}{e^2}\left[\frac{1}{4} F^2_{\mu\nu} + \frac{1}{2\xi} (\partial_\mu A_\mu)^2 + \frac{1}{2} m_\gamma^2 A^2_\mu + 
\overline{\psi}(x)\left(\gamma_\mu D_\mu + m\right)\psi(x)\right]\,,
\label{eq:qedm}
\end{equation}
where $D_\mu$ is the covariant derivative, $F_{\mu\nu}$ is the photon 
field tensor, and $m$ is the quark mass for a single flavour theory.
QED$_{\textrm{M}}$~features interesting properties: it is renormalisable, 
it is local, and it is amenable to numerical simulations with a minimal 
modification with respect to existing codes.

The tree-level photon propagator in a generic R$_\xi$ gauge is given by 
\begin{equation} 
D_{\mu\nu}(x) = \frac{1}{V}\sum_{q_\mu} \rm{e}^{iqx}\frac{1}{q^2+m_\gamma^2}\left[\delta_{\mu\nu} - 
\frac{q_\mu q_\nu(1-\xi)}{q^2 + \xi m_\gamma^2}\right]\,.
\label{eq:photon}
\end{equation}
One can immediately see that for $q_\mu=0$ the propagator is given 
by $\delta_{\mu\nu}/(m_\gamma^2V)$, thus the zero mode contribution diverges 
in the limit $m_\gamma \rightarrow 0$ at fixed finite volume.
A standard perturbative expansion in a finite volume 
would fail in this situation, because it treats all the modes on the same footing.
This is another aspect of the fact that, in the limit $m_\gamma \rightarrow 0$,
the zero modes do not appear in the Gaussian weight of the functional integral.
The role of the photon mass term is to provide a dynamical regulator of the zero modes.
The solution to this problem is to reorder the perturbative expansion including 
an arbitrary number of zero-mode propagators. 
To achieve this goal one treats the zero modes in the functional integral exactly,
and the non-zero modes as perturbative corrections.
To study the limit of $m_\gamma \rightarrow 0$ at fixed volume, it is thus necessary to 
treat the zero modes exactly.

To simplify the discussion below, and 
possibly introduce a new computational tool, we decompose the photon field into 
a zero mode contribution, $B_\mu$, and a non-zero mode fluctuation, $q_\mu(x)$
\begin{equation}
A_\mu(x) = B_\mu + q_\mu(x)\,.
\label{eq:decomposition}
\end{equation} 
The zero mode field, $B_\mu$, is constant in coordinate space, while the Fourier
decomposition of $q_\mu(x)$ does not contain the zero momentum term, i.e. 
$q_\mu(x) = \frac{1}{L^3 T}\sum_{k_\mu \neq 0} \rm{e}^{ikx} \widetilde{A}_\mu(k)$.
If $m_\gamma=0$, the zero (constant) modes are unconstrained by the gauge fixing procedure
and the theory, as a consequence of the LGT, is still invariant under a shift 
of the zero mode field~\cite{Blum:2007cy}
\begin{equation} 
B_\mu \rightarrow B_\mu +2 \pi \frac{r_\mu}{L_\mu}\,. 
\end{equation} 
This redundancy in the functional integral over the zero-mode fields 
cannot be eliminated by the local gauge fixing procedure, 
but rather by restricting the integration domain to $0 \le B_\mu \le \frac{2 \pi}{L_\mu}$.
If $m_\gamma=0$, the lattice theory in Eq.~\eqref{eq:qedm} is invariant under LGT,
thus for a generic fermionic correlation function there are $2$ possibilities: 
\begin{itemize}
    \item The fermionic correlation function is invariant under LGT (e.g. neutral particles). 
    In this case we can restrict the integration over the zero modes 
    in the domain $0 \le B_\mu \le \frac{2 \pi}{L_\mu}$.
    \item The fermionic correlation function is not invariant under LGT (e.g. charged particles).  
    In this case correlation functions vanish due to the LGT symmetry of the theory. We could insist in 
    defining the correlation functions restricting the integration domains of the zero modes in the 
    domain $0 \le B_\mu \le \frac{2 \pi}{L_\mu}$. In this way though, we are potentially introducing
    a non-locality in the correlation function. 
    The fact that the correlation function is not invariant under
    LGT does not guarantee any more that a LGT will bring the zero mode into 
    the restricted domain.
\end{itemize}
In the case of QED$_{\textrm{M}}$~with $m_\gamma \neq 0$ LGT are not a symmetry of the theory anymore.
This effectively allows us to calculate correlation 
functions with charged particles, with zero modes that are free to fluctuate
because we cannot use LGT anymore to restrict our integration domain.

We conclude that if we integrate the zero modes non-perturbatively in QED$_{\textrm{M}}$, we should not 
restrict our integration domain, because the photon mass regulates 
the fluctuations of the zero modes dynamically. 
Moreover we expect that correlation functions of charged correlators 
will vanish as we send the photon mass to zero, $m_\gamma \rightarrow 0$, in a finite volume $V$,
because in that limit we recover the symmetry under LGT.

The photon mass not only breaks LGT in a local way, but it also provides a mass gap to the theory.
The theory now possesses $2$ infrared (IR) cutoffs, $m_\gamma$ and $L$,
and the order in which these cutoffs are removed is important:
first one should perform the infinite volume limit, $L \rightarrow \infty$,
at fixed photon mass and then send the photon mass to zero, $m_\gamma \rightarrow 0$.

\section{Finite volume corrections}
\label{sec:fve}

To scrutinise whether QED$_{\textrm{M}}$~is a suitable lattice
QED formulation for precision physics, we need to address
finite size effects (FSE) specific to this action and 
understand the role played by the zero modes.

In Sec.~\ref{sec:qedm} we have argued that to study FSE of QED$_{\textrm{M}}$~
when lowering the photon mass, eventually the zero modes have to be
treated exactly. We introduce two distinct regimes
to analyse finite volume effects. The first regime
treats the zero modes of the photon on the same footing as the other modes.
In this regime the finite volume corrections are a consequence of the interaction,
mediated by photon fields, between copies of
matter particles located at distances multiples of $|L_\mu|$ between each other.
The interaction, mediated by massive photons, produces exponentially
suppressed finite volume corrections. This is somehow expected
and it is one of the main reasons to introduce QED$_{\textrm{M}}$.

Finite size corrections are calculated from the self-energy diagrams of matter
fields, of mass $M$, with the exchange of a virtual photon.  The electromagnetic
interaction with coupling $\alpha_{\rm{em}} = e^2/4\pi$, shifts the QCD mass of the matter
field by an amount $\Delta M = M(\alpha_{\rm{em}}) - M(\alpha_{\rm{em}}=0)$ that can be evaluated in
a finite volume, $V=L^3 \times T$, and in infinite volume.  The ultraviolet
divergences are the same in infinite and finite volume and they cancel out in
the subtraction $\delta_L M = \Delta M(L) - \Delta M$, where we consider the
time extent very large, i.e. $T \gg L$.  The calculation of the self-energy in
finite and infinite volume gives us the leading contribution proportional to
$m_\gamma$ \begin{equation} \delta_L M^{LO} = \pi \alpha_{\rm{em}} Q^2 2 m_\gamma
\mathcal{J}_1(m_\gamma L)\,,
\label{eq:FVLO}
\end{equation}
and the next-to-leading proportional to $m_\gamma^2/M$
\begin{equation}
\delta_L M^{NLO} = \pi \alpha_{\rm{em}} Q^2 m_\gamma \frac{m_\gamma}{M} \left[2 \mathcal{J}_{1/2}(m_\gamma L)
+  \mathcal{J}_{3/2}(m_\gamma L)\right]\,,
\label{eq:FVNLO}
\end{equation}
where $Q$ is the charge of the particle in units of $e$,
and $\mathcal{J}_n(z)$ are linear combinations of modified Bessel functions
\begin{equation}
\mathcal{J}_n(z) = \frac{1}{2^{n+{1 \over 2}} \pi^{{3 \over 2}}\Gamma(n)} \sum_{\mathbf\nu}
\frac{K_{{3 \over 2}-n}(z |\mathbf\nu| )}{(z | \mathbf\nu|)^{{3 \over 2}-n}}\,.
\end{equation}
In comparison with Ref.~\cite{Endres:2015gda}, this expression
does not include the contributions from the zero modes,
which we discuss separately in the next section.

\subsection{Zero modes}
\label{ssec:zero}

For the calculation of the zero-mode contributions to the finite size effects
we introduce a new computational tool and regime.
We call the region in parameter space where the zero modes of the photons
cannot be treated perturbatively the $\epsilon_\gamma$ regime.
In this regime we treat the zero modes of the photon field exactly in the
functional integral and we expand the non-zero mode contributions perturbatively.
In Eq.~\eqref{eq:decomposition} we have decomposed the photon field in a
constant field, $B_\mu$, and the fluctuations, $q_\mu(x)$, that do not contain zero modes.
If we rewrite the QED$_{\textrm{M}}$~action in Eq.~\eqref{eq:qedm} and,
as leading contribution, we set $q_\mu(x) = 0$ we obtain
\begin{equation}
S_{\rm{M}} = \frac{1}{2} \frac{m_\gamma^2 V}{e^2} B^2 + a^4 \sum_x
\overline{\psi}\left\{\gamma_\mu \left[\partial_\mu + i B_\mu\right] + M \right\} \psi\,,
\label{eq:qedm2}
\end{equation}
where we have removed all terms containing the non-zero modes fluctuations.

We notice that the zero modes contribute in two ways: a shift in the momentum of the
matter propagator from the matter part of the action,
and a Gaussian measure factor from the gauge part of the action.
We start by calculating the partition function
\begin{equation}
{\mathcal Z}_0 = \int {\mathcal D}[B_\mu,\psi,\overline{\psi}] \rm{e}^{-\frac{1}{2} \frac{m_\gamma^2 V}{e^2} B^2}
\rm{e}^{-\int d^4 x \overline{\psi}\left\{\gamma_\mu \left[\partial_\mu + i B_\mu\right] + M \right\} \psi}\,,
\end{equation}
where we have assumed that $q_\mu(x)=0$ and we consider the continuum formulation of the theory.
For the time being we neglect the functional integral on the fermions consistently with
the fact that in our numerical simulations we do not include QED fermion loops.
Neglecting the QED quark determinant we obtain
\begin{equation}
{\mathcal Z}_0 = \left(\frac{2 e^2 \pi}{m_\gamma^2 V} \right)^2\,.
\end{equation}
The next step is to calculate the two-point functions. Projecting to zero spatial momentum
\begin{equation}
C(x_4;{\bf p}=0) = \int d^3x~\left\langle \psi(x) \overline{\psi}(0)\right\rangle = \frac{1}{{\mathcal Z}_0} \int {\mathcal D}[B_\mu]
\rm{e}^{-\frac{1}{2} \frac{m_\gamma^2 V}{e^2} B^2}
\frac{1}{T} \sum_{k_4}\rm{e}^{ik_4x_4}\frac{-i \gamma_4 k_4 - i \slashed{B} + M}{\left(k_4 + B_4\right)^2 + \omega_B^2}\,,
\label{eq:corr_zm1}
\end{equation}
where the location of the pole is shifted because of the presence of the zero modes
\begin{equation}
\omega_B^2 = M^2 + |{\bf B}|^2\,.
\end{equation}
For simplicity, we neglect the Dirac structure in the numerator,
and consider only the propagator of a scalar particle.
The integrand
\begin{equation}
{\mathcal C}(x_4;{\bf p}=0;B) \rightarrow \frac{1}{T} \sum_{k_4}\rm{e}^{ik_4x_4}\frac{1}{\left(k_4 + B_4\right)^2 + \omega_B^2}\,,
\end{equation}
can be calculated using a Schwinger parametrisation.
The sum over $k_4$ can be done exactly and for large $T$ the leading term is given by
the infinite volume time-momentum representation
\begin{equation}
{\mathcal C}(x_4;{\bf p}=0;B) \simeq \rm{e}^{-i B_4 x_4}\frac{\rm{e}^{-\omega_B x_4}}{2 \omega_B}\,.
\end{equation}
We conclude that the leading finite size corrections
correspond to the integration over the zero modes of the
"infinite volume" propagator projected to zero momentum.
We can now finalise the calculation of
\begin{equation}
C(x_4;{\bf p}=0) = \frac{1}{{\mathcal Z}_0} \int {\mathcal D}[B_\mu]
\rm{e}^{-\frac{1}{2}\frac{m_\gamma^2 V}{e^2} B^2} \rm{e}^{-i B_4 x_4}\frac{\rm{e}^{-\omega_B x_4}}{2 \omega_B}\,.
\end{equation}
After few lines of algebra we obtain
\begin{equation}
C(x_4;{\bf p}=0) = \rm{e}^{-\zeta x_4^2} \rm{e}^{-M x_4} {\mathcal I}(x_4,M,m_\gamma,V)\,,
\label{eq:corr1}
\end{equation}
where we defined the zero mode parameter $\zeta$
\begin{equation}
\zeta = \frac{e^2}{2m_\gamma^2 V},
\label{eq:zeromode}
\end{equation}
and the integral ${\mathcal I}$ is given by
\begin{equation}
{\mathcal I}(x_4,M,m_\gamma,V) = \frac{1}{\overline{\mathcal{Z}}_0} \int d^3B~
\frac{\rm{e}^{-\frac{1}{2}\frac{m_\gamma^2 V}{e^2} |{\bf B}|^2}\rm{e}^{-(\omega_B-M)x_4}}{2 \omega_B}\,,
\label{eq:I_spatial_zm}
\end{equation}
where
\begin{equation}
\overline{\mathcal{Z}}_0 = \int d^3 B~\rm{e}^{-\frac{1}{2}\frac{m_\gamma^2 V}{e^2} |{\bf B}|^2}\,.
\end{equation}
We obtain the result of Ref.~\cite{Endres:2015gda} showing that the leading exponential behaviour
of the correlation function is modified by the zero modes, inducing a linear rise in $x_4$
in the effective mass proportional to $\zeta$.
This correction is a finite size effect and it does not imply the lack of a transfer matrix in the theory.
The lattice action is ultra-local and it would not be difficult to construct an explicit expression
for the transfer matrix. There are other examples, like the $\epsilon$ regime of
chiral perturbation theory, where the Euclidean time dependence of correlation
functions is not a sum of exponential functions and still
the theory possesses a transfer matrix~\cite{Shindler:2009ri}.
\begin{figure}
    \centering
    \includegraphics[width=.65\textwidth]{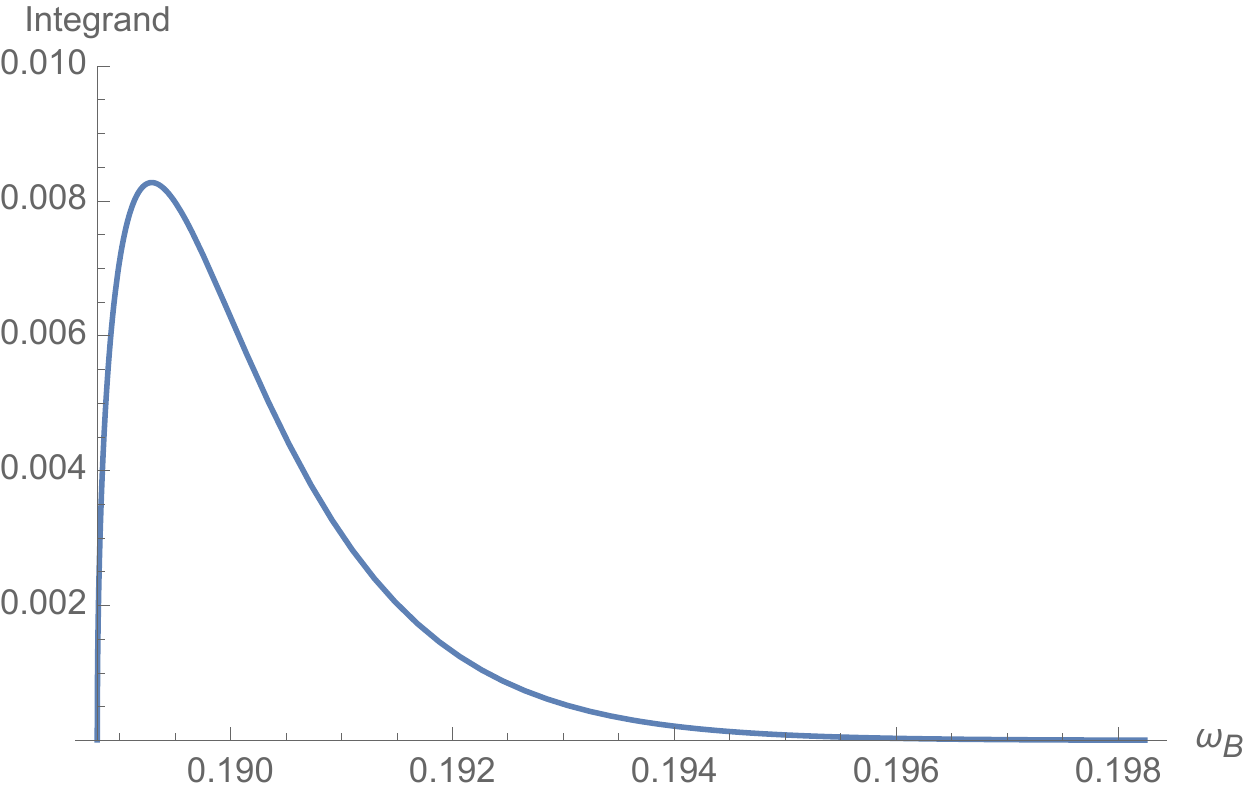}
    \caption{Integrand, Eq.~\eqref{eq:integrand}, for $am_\gamma=0.0236$, $m_\gamma/M=1/8$
    and $(L/a)^3 \times T/a = 24^3\times 64$.}
    \label{fig:integrand}
\end{figure}
We now want to scrutinise the contribution of the spatial zero modes in Eq.~\eqref{eq:I_spatial_zm}.
The integral ${\mathcal I}$ can be simplified using polar coordinates
\begin{equation}
{\mathcal I}(x_4,M,m_\gamma,V) = 2\pi \left(\frac{m_\gamma^2V}{2 e^2 \pi}\right)^{3/2}\rm{e}^{Mx_4}\rm{e}^{\frac{m_\gamma^2 M^2 V}{2 e^2}}
\int_M^\infty d\omega_B~\sqrt{\omega_B^2 - M^2} \rm{e}^{-\frac{1}{2}\frac{m_\gamma^2 V \omega_B^2}{e^2}}\rm{e}^{-\omega_B x_4}\,.
\label{eq:int_zmsp}
\end{equation}
In Fig.~\ref{fig:integrand} we plot the integrand
\begin{equation}
\sqrt{\omega_B^2 - M^2} \rm{e}^{-\frac{1}{2}\frac{m_\gamma^2 V (\omega_B^2-M^2)}{e^2}}\rm{e}^{-(\omega_B-M) x_4}\,,
\label{eq:integrand}
\end{equation}
as a function of $\omega_B$ for the most extreme value of the simulation parameters described below, i.e. $m_\gamma/M = 1/8$ and $V/a^4=24^3 \times 64$.
The peak is slightly shifted from $M$ and when increasing the volume both the location of the peak
and its width shrink to zero. The integrand itself vanishes when $V \rightarrow \infty$.
The analytical estimate of the integral can be done with a saddle point approximation.
To avoid the zero in the leading term, we include the sub-leading terms in $1/V$
in the approximation.
We rewrite
\begin{equation}
{\mathcal I}(x_4,M,m_\gamma,V) =
2\pi \left(\frac{m_\gamma^2V}{2 e^2 \pi}\right)^{3/2}
\int_M^\infty d\omega_B~\rm{e}^{V\left[-\frac{1}{2}\frac{m_\gamma^2 (\omega_B^2-M^2)}{g^2}
-\frac{(\omega_B-M)}{V} x_4 + \frac{1}{2V}\log\left(\omega_B^2-M^2)\right)\right]}\,.
\end{equation}
In the limit $V \rightarrow \infty$ we can now use the saddle point approximation including the
second and third exponentials that are sub-leading.
Using the usual machinery and evaluating the second derivative of the exponent
we find, up to higher order corrections in $1/V$
\begin{equation}
{\mathcal I}(x_4,M,m_\gamma,V) =
\left(\frac{8 \pi \alpha_{\rm{em}} m_\gamma^2V}{\rm{e} M^2}\right)^{1/2}
\rm{e}^{-\frac{\zeta}{M} x_4} \,,
\label{eq:mcI}
\end{equation}
where the saddle point is given by
\begin{equation}
\overline{\omega}_B = M\left(1+ \frac{\zeta}{M^2}\right)\,.
\end{equation}
To avoid confusion we denote the electric charge in terms of $\alpha_{\rm{em}} = e^2/4 \pi$,
while the $\rm{e}$ in Eq.~\eqref{eq:mcI} and in the following denotes Euler's number.
The correlation function in Eq.~\eqref{eq:corr1} now reads
\begin{equation}
C(x_4;{\bf p}=0) = \left(\frac{8 \pi \alpha_{\rm{em}} m_\gamma^2V}{\rm{e} M^2}\right)^{1/2}\rm{e}^{-\zeta x_4^2}
\rm{e}^{-M(1+\frac{\zeta}{M^2}) x_4}\,,
\end{equation}
thus the mass extracted from an effective mass analysis will contain two types of finite size effects
\begin{equation}
m_{\textrm{eff}} = -\frac{d}{dx_4}\log C(x_4;{\bf p}=0) = 2 \zeta x_4 + M\left(1+\frac{\zeta}{M^2}\right)\,,
\end{equation}
coming from the temporal zero modes, modifying the Euclidean time dependence
of the correlation function, and from spatial zero modes shifting the mass.
We notice, that as predicted in Sec.~\ref{sec:qedm} the correlation function
vanishes if we send the photon mass to zero at fixed finite volume.

For our ensemble at $M=M_\pi \simeq 310$ MeV, $(L/a)^3 \times T/a = 24^3\times
64$ and $m_\gamma/M_\pi=1/8,1/4$, which are the worse possible cases together
with the lightest photon mass $m_\gamma/M_\pi=1/8$ at the largest volume
$(L/a)^3 \times T/a = 48^3\times 64$, we obtain finite size effects relative to
the matter field mass of $0.024\%$, $0.006\%$ and $0.003\%$, respectively.  If
the QED corrections are at the percent level, these finite size corrections
contribute a $2.4\%$ correction in the worse case. As discussed in
Sec.~\ref{sec:fv}, the smallest photon mass will not be included in the final
analysis.  We conclude that once the temporal zero modes corrections are exactly
subtracted from the effective mass analysis the remaining effects stemming from
the spatial zero modes can be safely neglected, possibly with the exception of
the case $(L/a)^3 \times T/a = 24^3\times 64$ and $m_\gamma/M_\pi=1/8$.

\subsection{Dispersion relation}

A final remark concerns the contribution of the zero modes to the energy
spectrum at non-zero spatial momentum.  In Ref.~\cite{Patella:2017fgk} it was
pointed out that finite volume corrections for QED$_{\textrm{M}}$~can cause a suppression of
the dependence on the momentum in the effective energy.
\begin{figure}
  \centering
  \includegraphics[width=.48\textwidth]{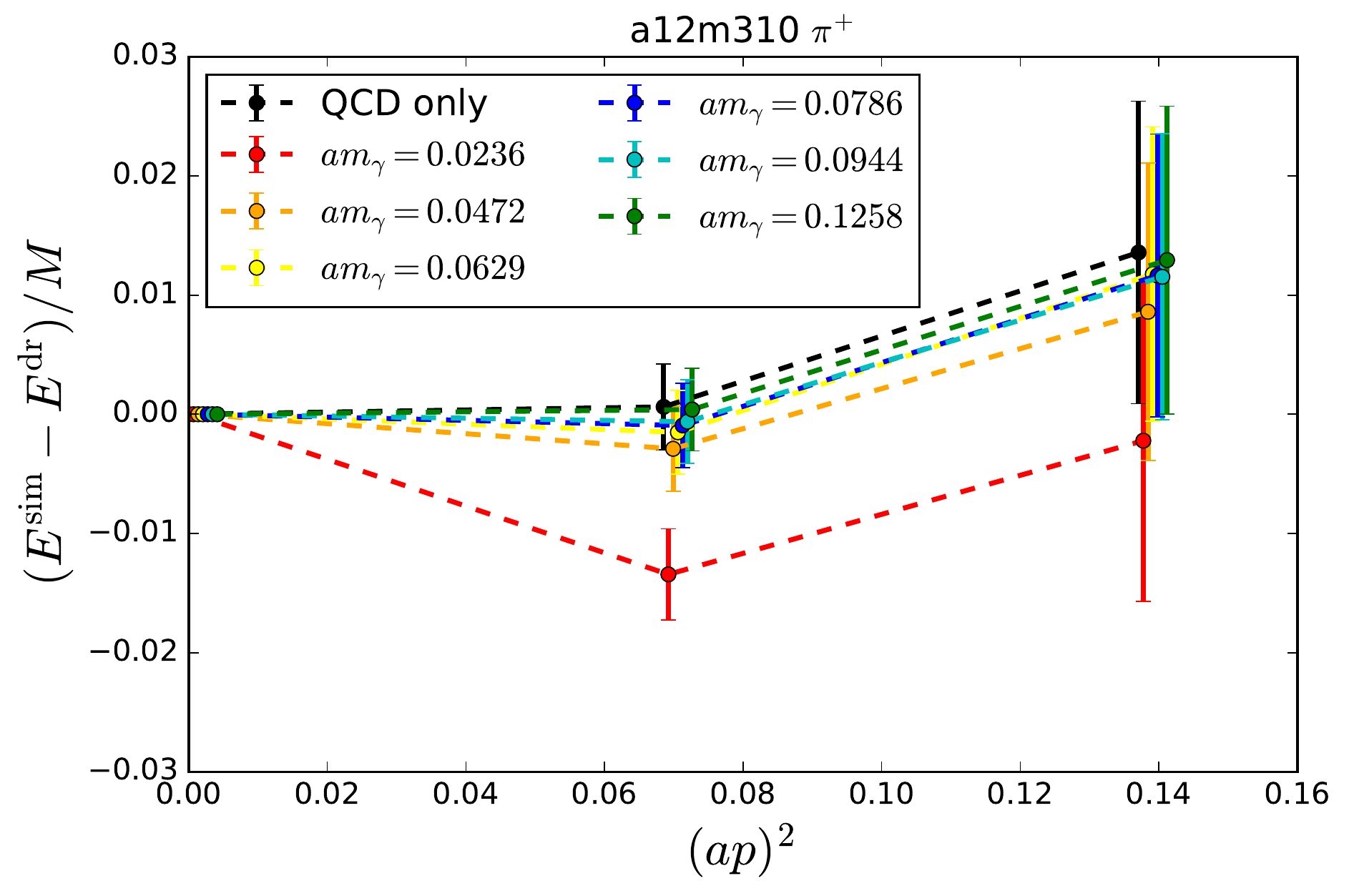}
  \includegraphics[width=.48\textwidth]{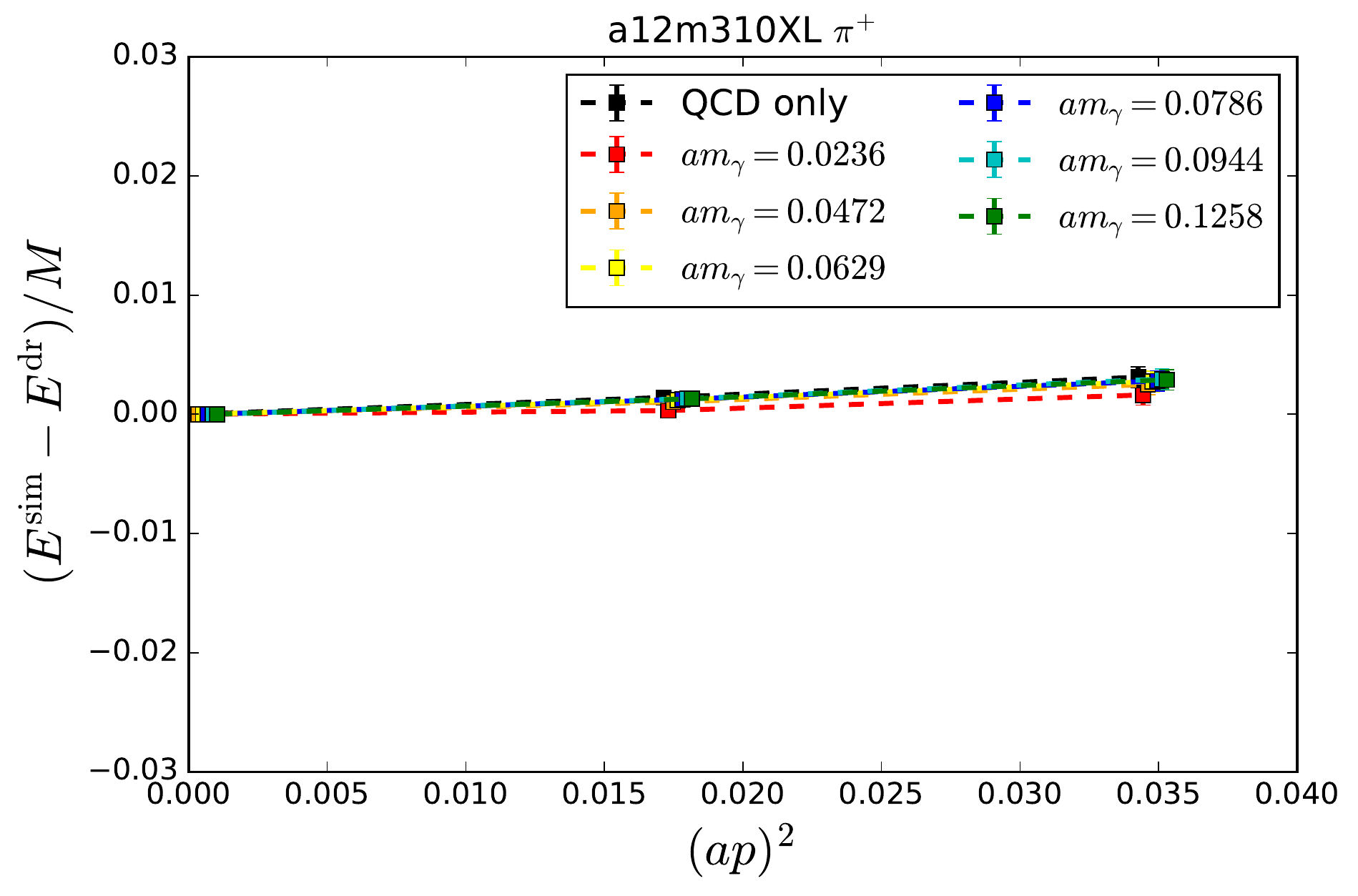}
  \caption{Relative discrepancy between the expected lattice dispersion relation
    and the computed energy for the charged pion as a function of momentum for
    the various simulated photon masses on the {\tt a12m310} (left) and the {\tt
      a12m310XL} (right) ensembles.  To ease the comparison we use the same
    scale in the y-axis.}
  \label{fig:disp}
\end{figure}
In other words if the photon mass is too small at fixed finite volume, the
dispersion relation can be violated. We determined the energy of the charged
pion as a function of the momentum of the pion, $E^{\rm{sim}}(p^2)$ for several
photon masses (see Sec.~\ref{sec:dataanalysis} for details of the numerical
analysis).  We then compared the results from the simulations to the expected
lattice dispersion relation
\begin{equation}
  \sinh\left( \frac{a E^{\textrm{dr}}}{2}\right) = \sinh \left(\frac{a
    M}{2}\right) + \sum_{i=1}^3 \sin^2\left(\frac{a p_i}{2} \right)\,.
\end{equation}
In Fig.~\ref{fig:disp} we show
$\frac{E^{\rm{sim}}(p^2) - E^{\textrm{dr}}}{M}$ as a function of
$p^2$. Different colours show results with different photon masses; the left
plot corresponds to the small volume, while the right plot the larger volume. We
observe that, beside the smallest photon mass in the small volume, all the
energies computed satisfy the expected dispersion relation. As a check of the
procedure we also show the same quantity for pure QCD, obtaining the same type
of agreement.

We conclude that the dispersion relation is a useful quantity to monitor in
order to check the contribution of the photon zero modes. In our numerical
studies only the smallest photon mass on the ensemble with the smaller volume,
corresponding to $m_\gamma L=0.53$, deviates from the expected form of the
dispersion relation.

\section{Computational Set-Up}
Having laid the analytic ground work to assess the finite size effects we now
turn our attention to numerical tests. The main goal of this section is to
establish a region in parameter space, where the required limits~i.e. the
extrapolations to the infinite volume and subsequently to vanishing photon mass,
are well controlled. To this end, we compute the mass splittings due to QED
isospin breaking effects for two ensembles with identical parameters except the
spatial volume. On these ensembles we simulate a range of six different photon
masses given by $m_\gamma = \{1/8, 1/4, 1/3,5/12,1/2,2/3\} \times M_\pi$. If the
finite size effects are indeed well described, the simulations performed on the
different volumes need to produce compatible results once the infinite volume
extrapolation has been performed.

Numerically, we utilise the mixed-action set-up established by the CalLat
collaboration~\cite{Berkowitz:2017opd} using gauge field configurations with
$N_f=2+1+1$ dynamical highly improved (rooted) staggered quarks
(HISQ)~\cite{Follana:2006rc} in the sea which are gradient flowed prior to
inverting M\"obius domain wall fermions in the valence sector. We use the {\tt
  a12m310} (from MILC~\cite{MILC:2010pul}) and {\tt a12m310XL} (from
CalLat~\cite{Miller:2020evg}) ensembles. These ensembles have a lattice spacing
$a \approx 0.12$~fm, identical sea quark masses resulting in a pion mass of
approximately $310\,\mathrm{MeV}$ and 4-volumes of $L^3\times T/a^4 = 24^3\times
64$ and $48^3\times 64$. Our computations are performed in the electro-quenched
approximation using stochastic $U(1)$ background gauge fields which are fixed to
Landau gauge.

\section{Data Analysis \label{sec:dataanalysis}}
We generate two-point correlation functions $C_{\mathbb{P}^Q}^{SY}(t;m_\gamma)$
inducing the quantum numbers of a state $\mathbb{P}$ with charge $Q$. Here the
``S'' indicates that the source of the propagators is always Gaussian smeared,
whilst the sink can be point-like or smeared ($Y\in \{P,S\}$). Omitting the
backwards travelling contribution to the correlation functions, we write the
isospin symmetric correlation function as
\begin{equation}
  C^{SY}_{\bar{\mathbb{P}}}(t) = \sum_i \frac{\bar{Z}_i^S \bar{Z}_i^Y}{2\bar{M}_i} e^{-\bar{M}_it}.
\end{equation}

We express the masses and amplitudes in the presence of QED as the splittings $\delta Z_i$ and $\delta M_i$ to the isospin symmetric quantities, i.e.

\begin{equation}
  C^{SY}_{\mathbb{P}^Q}(t;m_\gamma) = \sum_i \frac{\left(\bar{Z}_i^S+\delta Z_i^S(m_\gamma)\right) \left(\bar{Z}_i^Y+\delta Z_i^Y(m_\gamma)\right) }{2\left(\bar{M}_i+\delta M_i(m_\gamma)\right)} e^{-\left(\bar{M}_i+\delta M_i(m_\gamma)\right)t-\zeta(m_\gamma)t^2},
\end{equation}
where $\zeta(m_\gamma)$ is the contribution from the zero-mode as defined in
\eqref{eq:zeromode}.  For light mesonic states we have to consider the backwards
travelling contributions given by a second term with $t \to (T-t)$. Note that
this can be simplified using
\begin{equation}
  \begin{aligned}
    e^{-mt} + e^{-m(T-t)} &= 2e^{- \frac{mT}{2}} \mathrm{cosh}\left[ m\left(t-\frac{T}{2}\right) \right],\\
    e^{-mt-\zeta t^2} + e^{-m(T-t)-\zeta(T-t)^2} &= 2e^{-\frac{(m+\zeta T)T}{2}} \mathrm{cosh}\left[\left(m+\zeta T\right)\left(t-\frac{T}{2}\right) \right] e^{-\zeta \left(t^2-tT\right) }. 
  \end{aligned}
  \label{eq:zeromodecors}
\end{equation}

Even though we have a large data set for the meson and baryon octets and the
baryon decouplet, here we are mostly interested in the sensitivity to the
infrared regularisation so in the following we will mostly focus on the lightest
states $\mathbb{P}^Q=\pi^\pm$, $\pi^0$, $K^\pm$ and $K^0$.

\subsection{Correlator analysis and zero mode}

We note that the term involving the zero mode $\zeta$ is analytically completely
determined. Prior to any correlation function analysis, we remove this by hand
by multiplying the relevant correlation functions by $e^{\zeta t^2}$ (if the
backwards propagating part is negligible) and $e^{\zeta (t^2-tT)}$
(otherwise). In the latter case, one can then fit the correlation function to
the typical functional form and obtain $m+\zeta T$ (see
\eqref{eq:zeromodecors}). Clearly $\zeta T$ can then be subtracted analytically.

\begin{figure}
  \includegraphics[width=\textwidth]{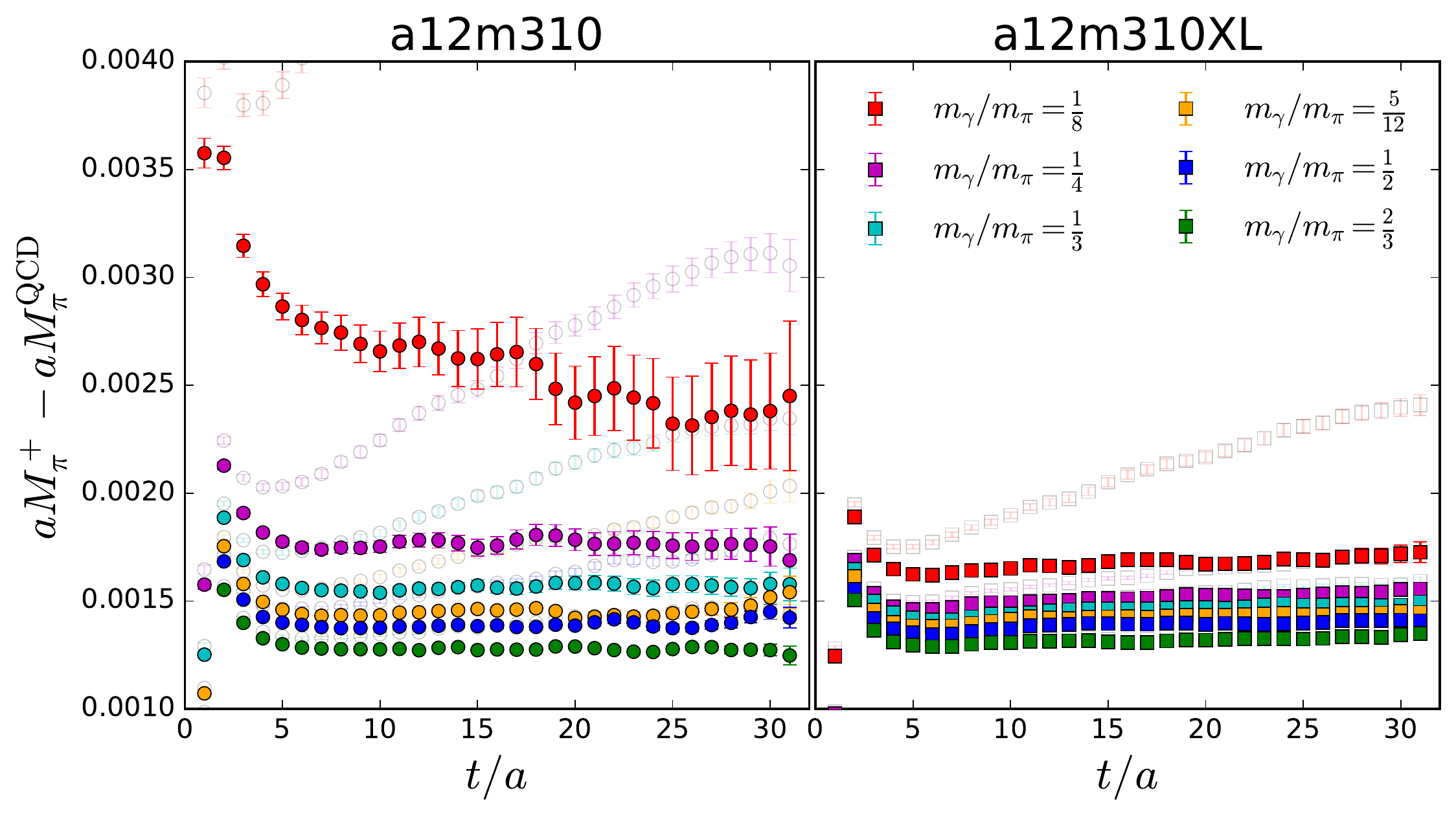}
  \caption{Effective mass splittings for charged pions. The correlated
    difference between effective masses in the presence of QED and the QCD
    effective mass is shown. The zero mode leads to a linear rise in the
    effective mass (faint symbols), which can be removed analytically (filled
    symbols).}
  \label{fig:zeromode}
\end{figure}

Figure \ref{fig:zeromode} shows the effective mass splitting for the charged
pion as a function of photon mass (different colours) and spatial volume (left
vs right panel). This is obtained by taking the correlated difference of the
effective masses from the correlation functions with QED to those using the QCD
only correlation function. The faint symbols show a linear rise stemming from
the zero mode. The closed symbols demonstrate that once the zero mode is
removed, one obtains very stable plateaus in the effective mass splitting. When
comparing the effective mass splittings between the {\tt a12m310} (left) and the
{\tt a12m310XL} (right) ensembles, we notice a significant dependence on the
volume. We re-iterate that these two simulations only differ in the doubled
spatial extents of the {\tt a12m310XL} ensemble.

\begin{figure}
  \centering
  \includegraphics[width=.6\textwidth]{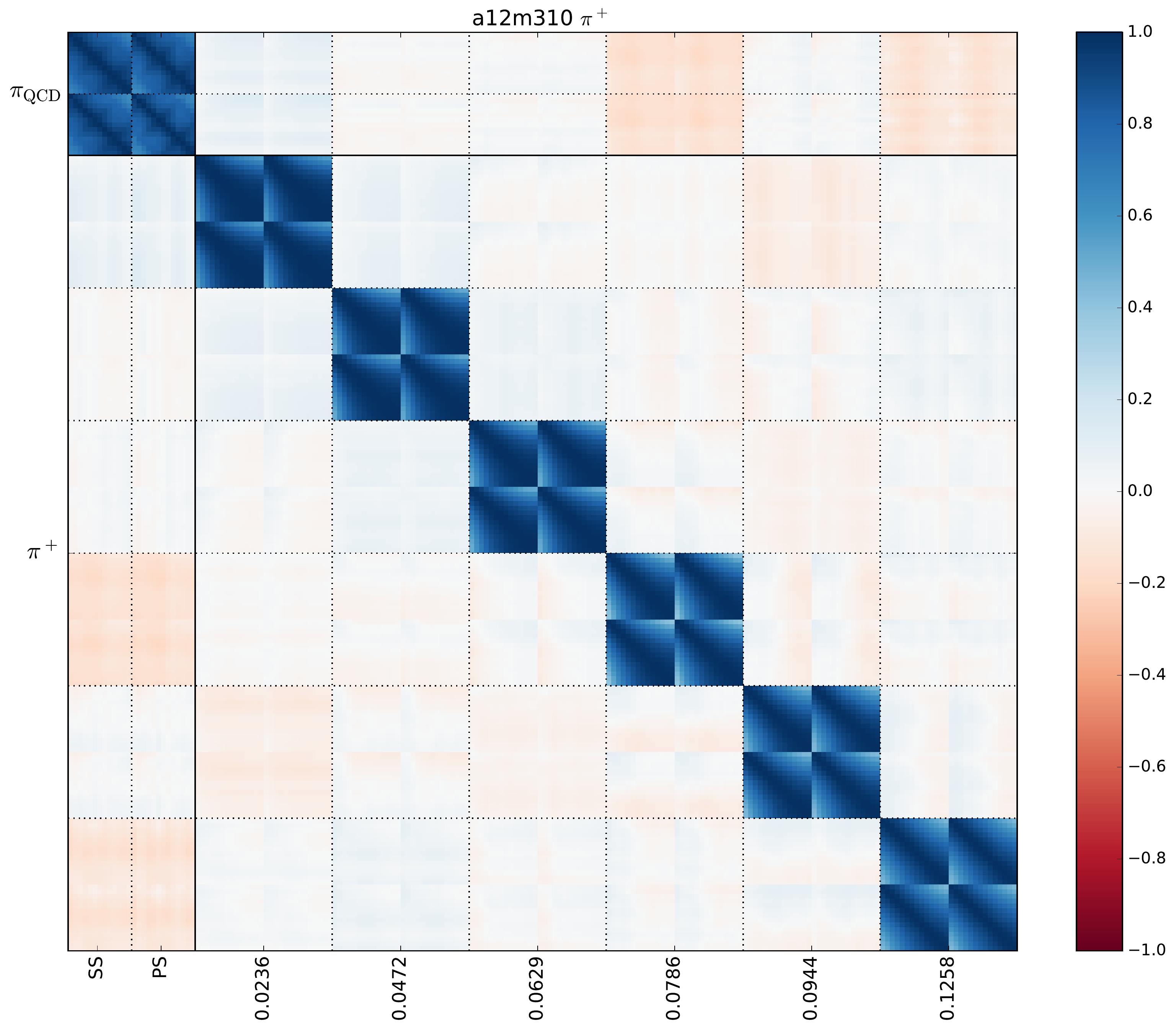}
  \caption{Correlation matrix entering the correlated fit to extract the QCD
    parameters and the QED induced splittings.}
  \label{fig:correlation}
\end{figure}

In order to extract mass splittings it is beneficial to consider the following
ratios of correlation functions
\begin{equation}
  \begin{aligned}
    R^{SY}_{\mathbb{P}^Q}(t;m_\gamma) &= \frac{e^{-\zeta \left(t^2+\frac{T^2}{2}-tT\right)} C^{SY}_{\mathbb{P}^Q}(t;m_\gamma)}{C^{SY}_{\bar{\mathbb{P}}}(t)} \\
    &= \frac{\sum_i \left(\bar{Z}_i^S+\delta Z_i^S \right) \left(\bar{Z}_i^Y+\delta Z_i^Y\right) \mathrm{cosh}\left[\left(\bar{M}_i+\delta M_i+\zeta T\right)\left(t-\frac{T}{2}\right)\right] e^{-(\bar{M}_i+\delta M_i) T/2}/\left(\bar{M}_i+\delta M_i\right)}{\sum_j \bar{Z}_j^S\bar{Z}_j^Y  \mathrm{cosh}\left[\bar{M}_j\left(t-\frac{T}{2}\right)\right] e^{-\bar{M}_j T/2}/\bar{M}_j}\\
  &\approx \left(1+\frac{\delta Z^{S}_0}{Z^{S}_0}+\frac{\delta Z^{Y}_0}{Z^{Y}_0} - \frac{\delta M_0}{\bar{M_0}}\right)  e^{-\delta M_0t} +\mathcal{O}(e^{-\Delta \bar{M}_1t},e^{-(\Delta \bar{M}_1 +\delta M_0)t},e^{-(\Delta \bar{M}_1 +\delta M_1)t},\delta^2),
  \end{aligned}
\end{equation}
where $\delta X_i$ refers to the QED-induced splittings of the i$^\mathrm{th}$
excited state of the observable $X$ and $\Delta X_i$ to splitting between the
i$^\mathrm{th}$ excited state and the ground state. In addition to being very
sensitive to the mass splitting, the high degree of correlation between the two
correlation functions that enter the ratio results in very favourable
statistical properties. This leads to the correlation matrix that enters the fit
to be close to block diagonal. An example of such a correlation matrix is shown
for the charged pion correlation functions on the {\tt a12m310} ensemble in
Figure \ref{fig:correlation}.

\begin{figure}
  \includegraphics[width=\textwidth]{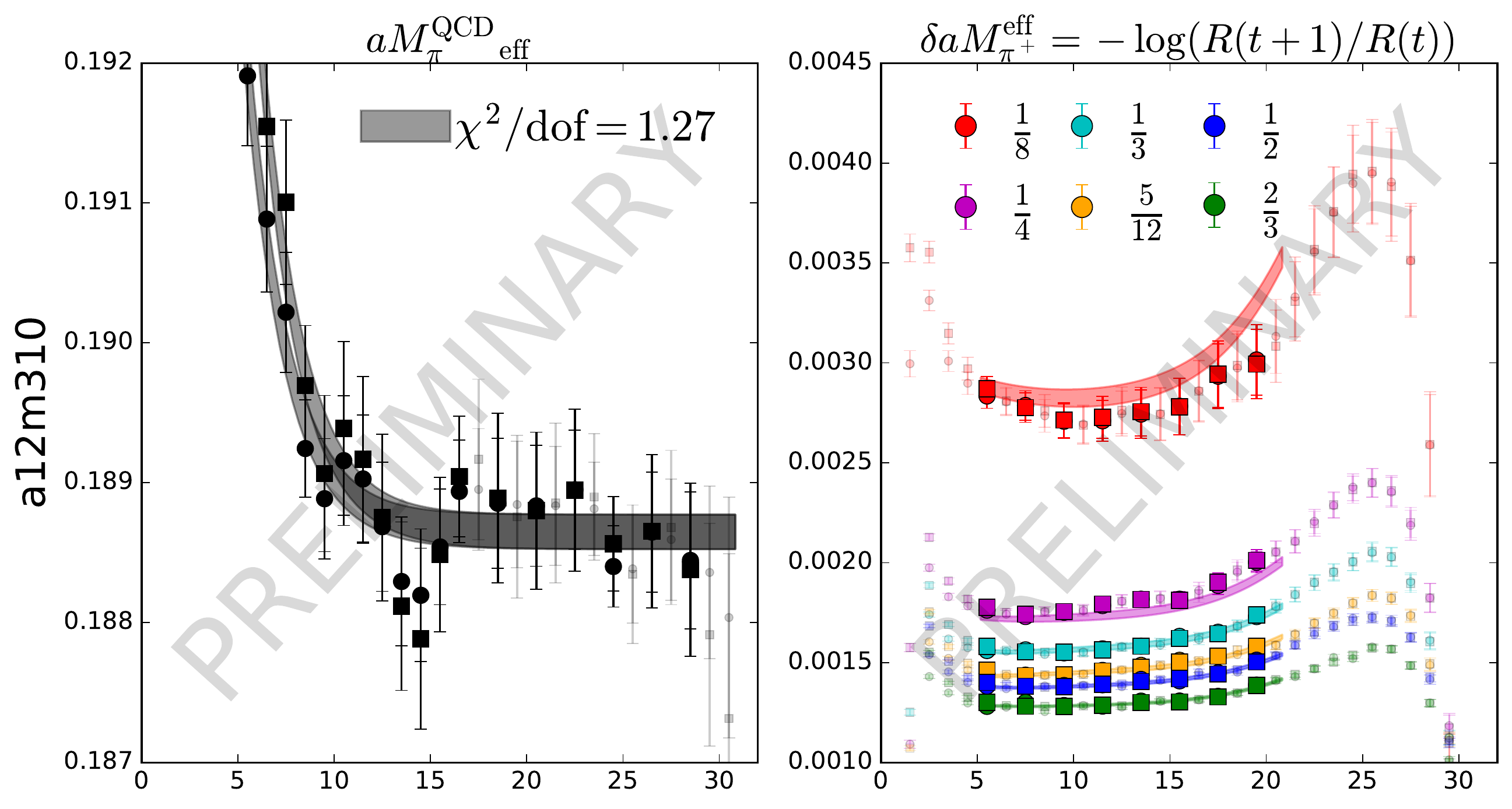}
  \caption{Fit results superimposed to the effective mass (left) and mass
    splittings (right) for the charged pion.}
  \label{fig:corfit}
\end{figure}

\begin{figure}
  \centering
  \includegraphics[width=.8\textwidth]{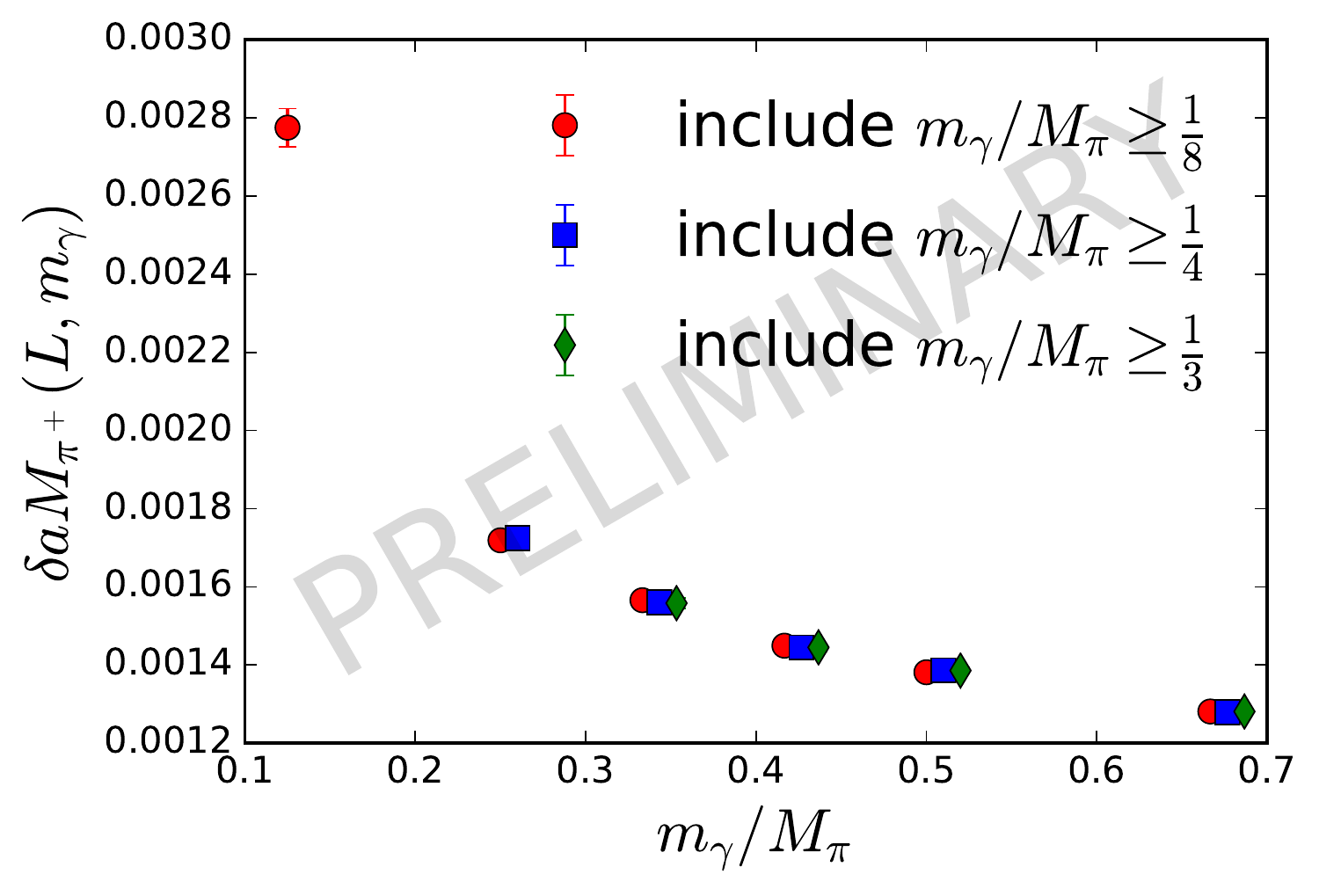}
  \caption{Extracted QED-induced mass splittings in the pion for a number of
    cuts to the smallest included photon mass on the {\tt a12m310} ensemble. The
    data points are slightly staggered along the horizontal axis for better
    visibility.}
  \label{fig:stability_pi}
\end{figure}

In order to obtain all required matrix elements, masses and their splittings, we
perform a simultaneous correlated fit to the isospin symmetric correlation
functions $C^{SY}_{\mathbb{P}^Q}(t;m_\gamma)$ and the ratios
$R^{SY}_{\mathbb{P}^Q}(t;m_\gamma)$ including the ground state and first excited
state throughout. An example projection to the effective masses of such a fit is
shown in Figure \ref{fig:corfit}. All 14 correlation functions (6 different
photon masses and the QCD only correlation function, each for two different
smearing choices) are well described by a single correlated fit. We assess the
stability of this fit by systematically varying the fit ranges as well as by
consecutively removing the data corresponding to the lightest photon masses from
the fit. We show the stability of the mass splittings to the isospin symmetric
pion mass in Figure \ref{fig:stability_pi}. We observe that removing the
lightest photon mass(es) from the fit does not change the remaining fit results,
indicating stability.

\subsection{Infinite volume limit \label{sec:fv}}
In order to make precise predictions with fully controlled systematic
uncertainties, one has to carefully investigate all required limits. Before
taking the vanishing photon mass limit, we need to correct the data for the
finite size effects and extrapolate to infinite volume. As discussed above,
given the infrared nature of the photon mass, the finite size effects are of
crucial importance. We generated data on two ensembles with identical parameters
other than their spatial volume, precisely for this reason, i.e. this study is
designed to test how accurately these corrections describe our data.  We test
this by removing the leading and next to leading order finite size corrections
from the data by applying the corrections detailed in \eqref{eq:FVLO} and
\eqref{eq:FVNLO}.  However, care must be taken since the zero-mode contributions
are assumed to be present in the development of the formalism for the finite
volume corrections, but the zero mode has already been removed by hand. We
follow the prescription described in Ref.~\cite{Endres:2015gda} and add and
subtract the zero mode in order to complete the sum in \eqref{eq:FVLO}. This
results in an additional finite volume correction $\delta_L^\mathrm{zm}$ defined
by
\begin{equation}
  \delta_L^\mathrm{zm} = -2\pi\alpha Q^2 m_\gamma \frac{1}{(m_\gamma L)^3}.
  \label{eq:FVzm}
\end{equation}

We stress that this is work in progress and likely only an effective
approximation for sufficiently large temporal extents. One concern is that the
zero mode contribution \eqref{eq:FVzm} does not vanish as $T \to \infty$, even
though the quantity $\zeta$ defined in \eqref{eq:zeromode} does. Given the
identical temporal extent of the two ensembles presented here, this contribution
is expected to be the same for both ensembles, thereby allowing us to
meaningfully compare the finite volume effects using the data at hand. We expect
to provide further analytical work and numerical evidence, taking the finite
four-volume into account in the near future.

\begin{figure}
  \centering
  \includegraphics[width=.47\textwidth]{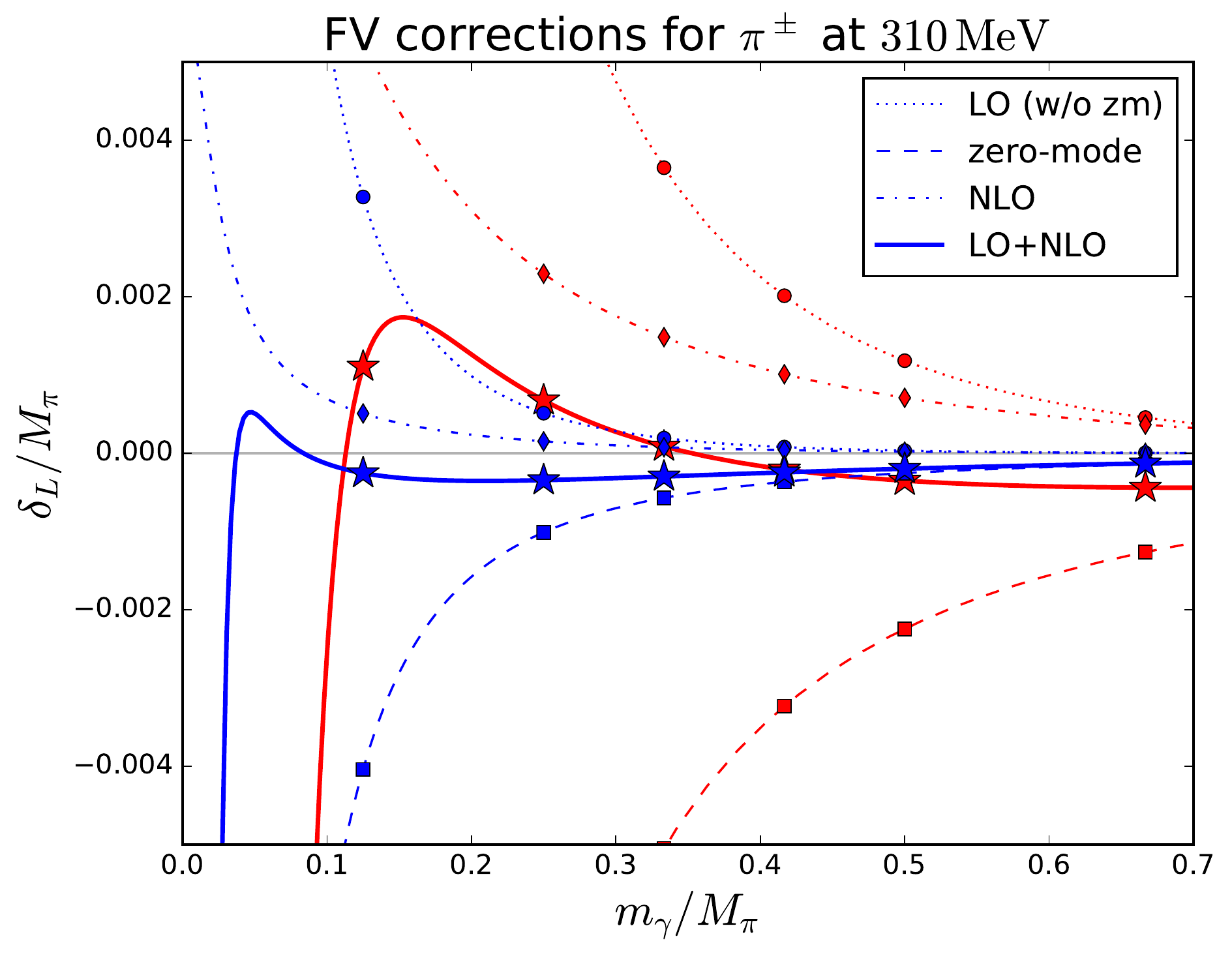}
  \includegraphics[width=.47\textwidth]{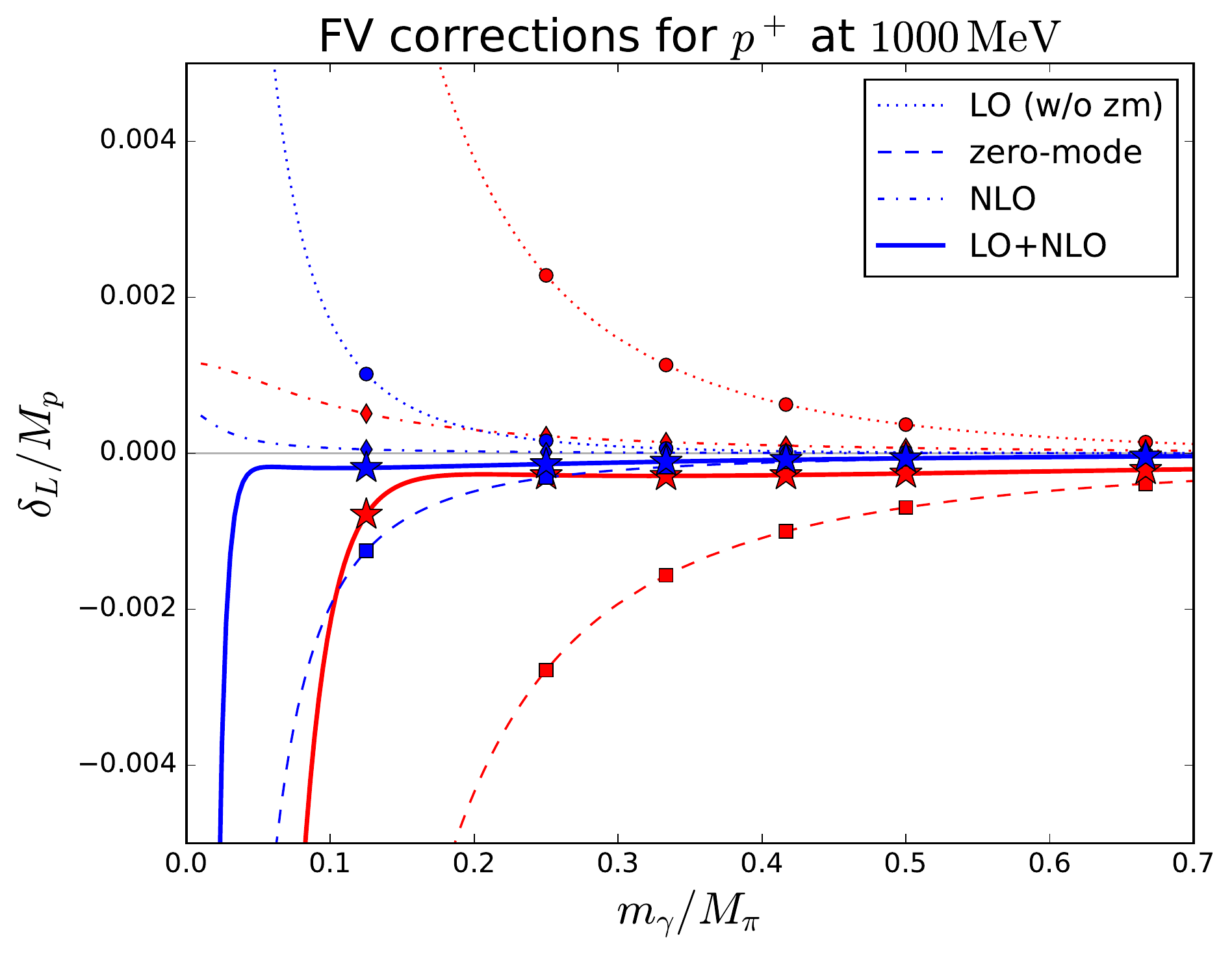}
  \caption{Individual contributions to the finite volume corrections as a
    function of $m_\gamma/M_\pi$. The red (blue) data represent the lattice
    parameters of the {\tt a12m310} ({\tt a12m310XL}) ensemble. The left panel
    shows the relative size for a pion mass of $310\,\mathrm{MeV}$, the right
    hand panel for a proton of $1000\,\mathrm{MeV}$. The symbols correspond to
    the simulated data points on these ensembles.}
  \label{fig:FVcontributions}
\end{figure}

Figure \ref{fig:FVcontributions} shows the various contributions that are
subtracted from the data to recover a good approximation for the infinite volume
limit. In particular $\delta_L^\mathrm{LO}$ (dotted lines),
$\delta_L^\mathrm{zm}$ (dashed lines) and $\delta_L^\mathrm{NLO}$ (dash-dotted
lines) are shown as well as their total (solid lines). The ensemble with the
smaller volume ({\tt a12m310}) is shown in red, whilst the larger volume result
({\tt a12m310XL}) is shown in blue. These contributions are evaluated for a pion
mass of $310\,\mathrm{MeV}$ and a proton mass of $1000\,\mathrm{MeV}$. As
expected the individual finite size effects are larger for the smaller
volume. Also in agreement with expectation, the finite size effects are more
pronounced for the pion than for a proton.

We observe that $\delta_L^\mathrm{zm}$ and $\delta_L^\mathrm{LO}$ are similar in
magnitude but have an opposite sign leading to large cancellations, causing the
sum of the contributions to change sign for very small photon masses. Clearly
the finite volume formulae are insufficient to accurately describe the
corrections for such very small photon masses. For this reason we remove the
smallest photon masses on the smaller volume in all cases and carefully study
the effect of removing further photon masses close to this regime.

\begin{figure}
  \centering
  \includegraphics[width=.47\textwidth]{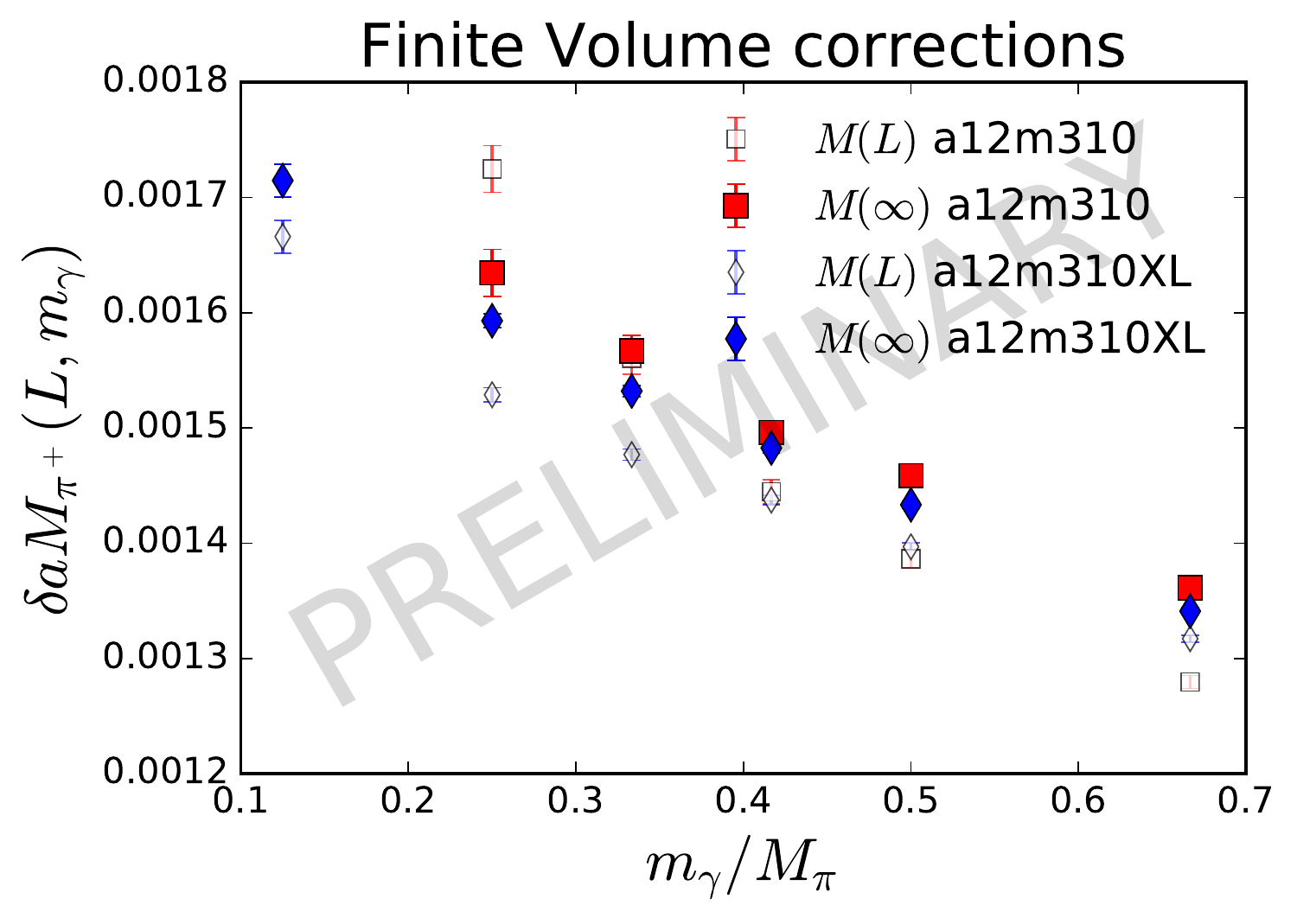}
  \includegraphics[width=.47\textwidth]{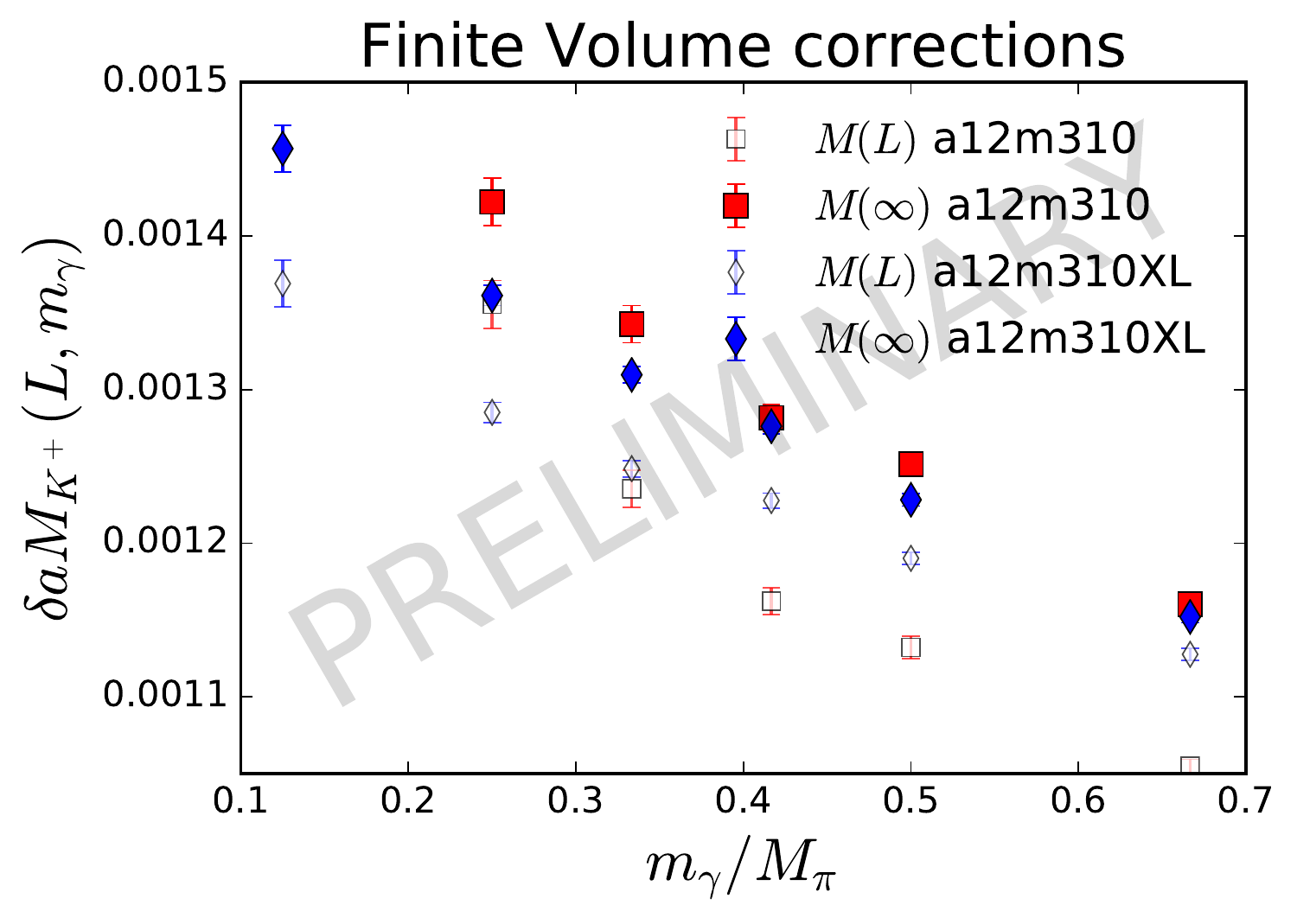}
  \caption{The effect of the finite volume corrections in the QED-induced mass
    splitting of the charged pion (left) and the charged kaon (right). Open
    symbols show the mass splittings extracted from the simulation, closed
    symbols the mass splittings after the finite volume corrections have been
    applied.}
  \label{fig:FVdata}
\end{figure}

Figure \ref{fig:FVdata} shows the mass splittings of the charged pion (left) and
the charged kaon (right) for the two simulated ensembles as a function of the
photon mass. The simulated data points prior to any finite size corrections are
shown as the open symbols, the mass splittings after the finite size corrections
have been applied are shown as the closed symbols. After the leading order and
next-to-leading order finite size effects have been removed, the data show good
agreement between the two ensembles, particularly for the larger photon masses.

\subsection{Vanishing photon mass limit}
Finally it remains to take the $m_\gamma \to 0$ limit to remove the second
infrared regulator. This is guided by an effective field theory
approach~\cite{Endres:2015gda}. The limit $m_\gamma \to 0$ is parameterised as
\begin{equation}
  M(0) = M(m_\gamma) - \Delta_\gamma M^{\mathrm{LO}} - \Delta_\gamma M^{\mathrm{NLO}} +\mathcal{O}\left(\frac{m_\gamma^3}{M^2}\right),
\end{equation}
where the leading order and next-to-leading order
corrections are given by
\begin{equation}
  \Delta_\gamma M^{\mathrm{LO}} = - \frac{\alpha}{2} Q^2 m_\gamma
\end{equation}
and
\begin{equation}
  \Delta_\gamma M^{\mathrm{NLO}} = \left(C\alpha - \frac{\alpha}{4\pi} Q^2\right) \frac{m_\gamma^2}{M}.
\end{equation}
Here $M(0)$ is the result for the mass (or mass splitting) at zero photon mass,
$Q$ is the charge of the hadron and $C$ is a low energy constant that can be
determined through a fit to the data.  This correction arises from
``hard-photon'' corrections between the quarks comprising the hadron.

\begin{figure}
  \centering
  \includegraphics[width=.47\textwidth]{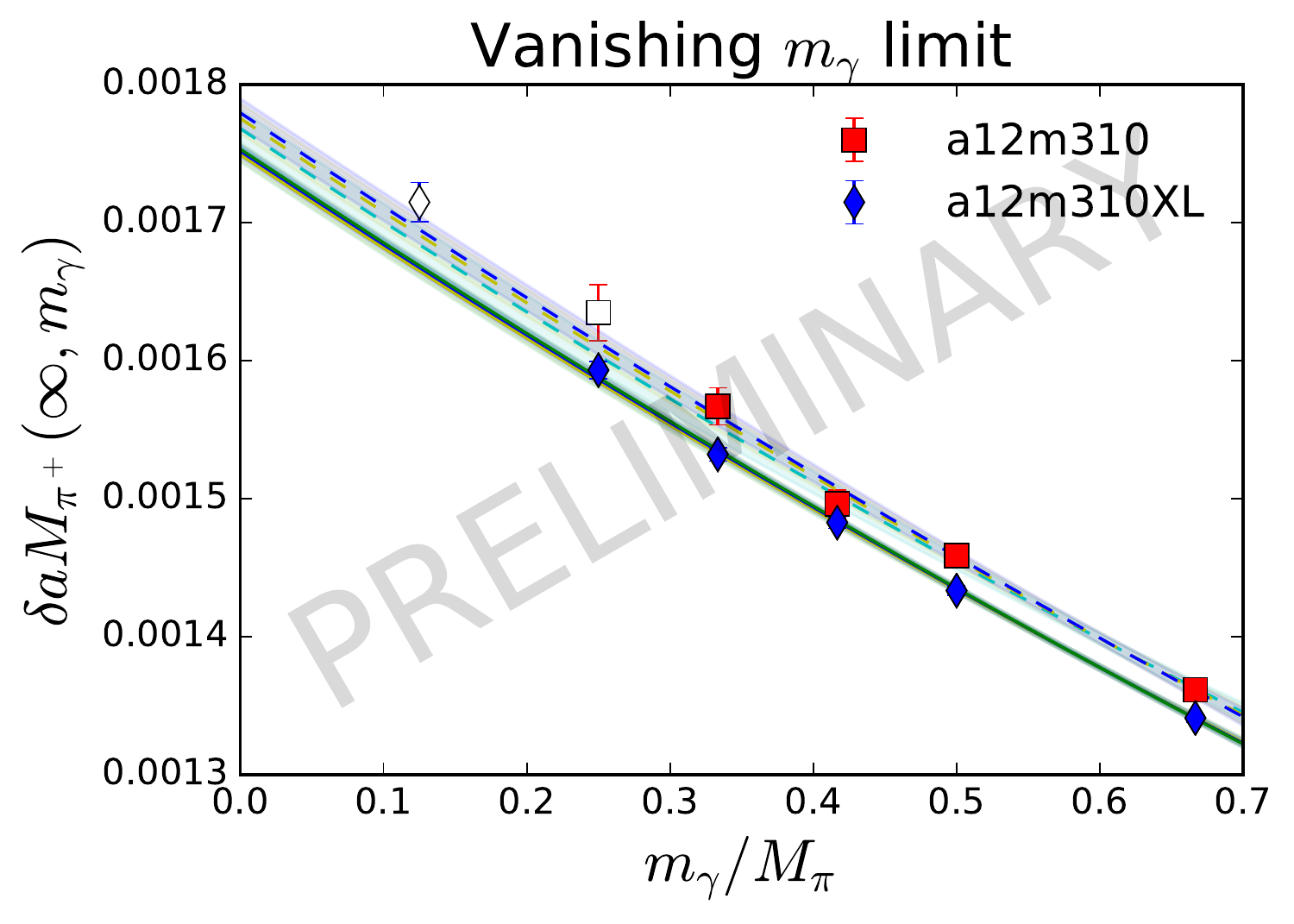}
  \includegraphics[width=.47\textwidth]{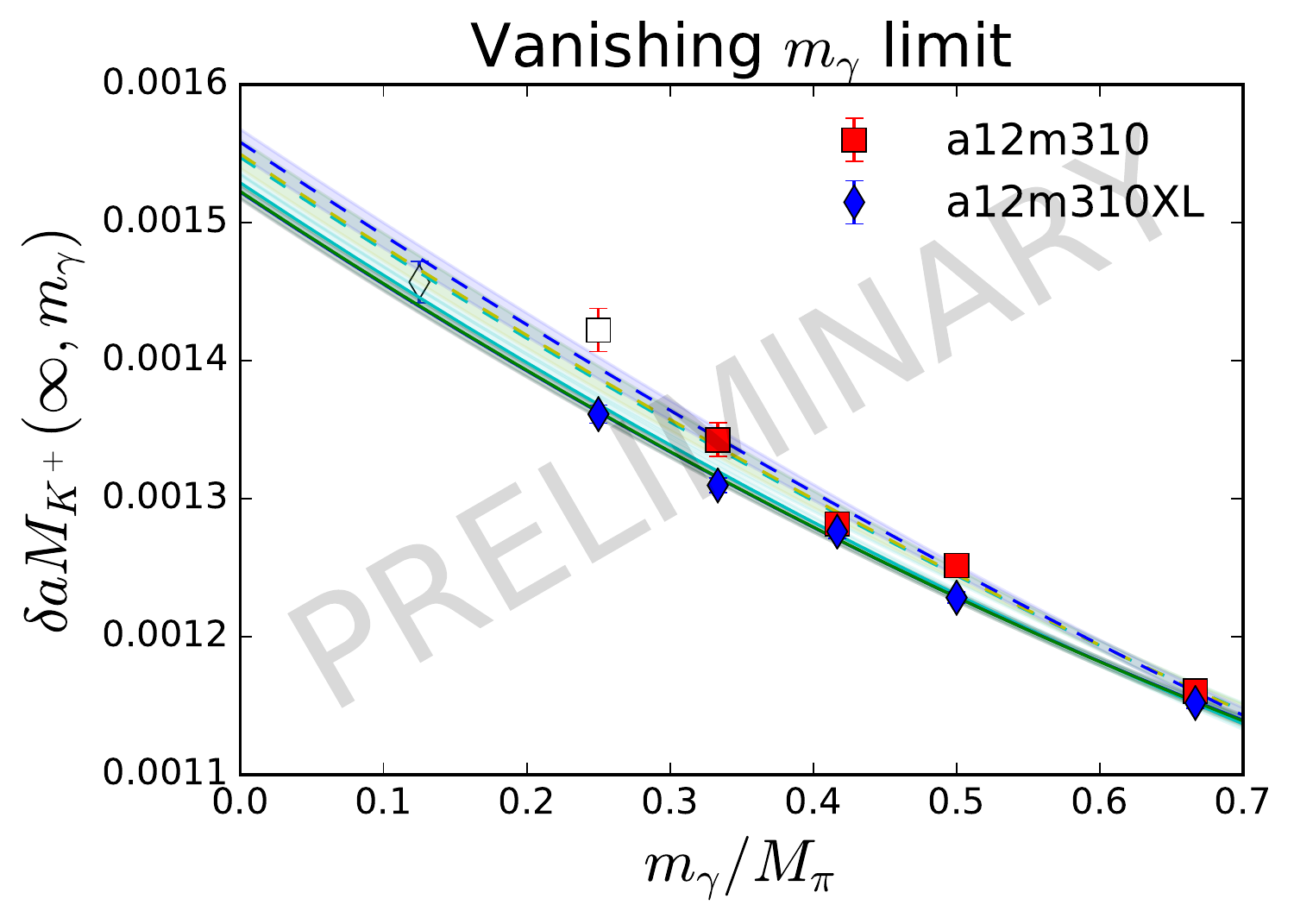}
  \caption{Examples of the vanishing photon mass extrapolation of the data
    obtained by the finite volume corrections for the case of the mass splitting
    of the charged pion (left) and the charged kaon (right). }
  \label{fig:mgamma}
\end{figure}
We perform the extrapolation as a fit to the different photon mass data
points. Figure \ref{fig:mgamma} shows the result for the mass splitting of the
charged pion (left) and the charged kaon (right) on the two ensembles. The data
points in these plots are the results from applying the finite size effects in
the previous section. The closed symbols correspond to data points entering the
chosen fit, the open symbols are excluded from the chosen fit. Additional
coloured lines correspond to alternative fits where consecutively more of the
lighter photon masses are removed, which all provide stable results.

Due to the size of the finite size effects we exclude the lightest photon mass
on the larger ensemble (red squares, {\tt a12m310XL}) and the lightest two
photon masses on the smaller ensemble (blue diamonds, {\tt a12m310}). We obtain
statistically compatible low energy constants $C$ from the two fits. This
preliminary analysis is based solely on statistical uncertainties, i.e. we have
not yet included any systematic error budget for choices made in the correlation
function fits or included estimates of higher order finite volume
corrections. The agreement at vanishing photon mass is very encouraging and
indicates that all the required limits are under control.

\section{Summary and Outlook}
We have presented numerical results using massive QED to compute QED effects in
precision lattice QCD simulations. We find a good statistical signal, allowing
us to determine QED induced splittings of masses and matrix elements though a
suitably chosen combined fit to correlation functions and ratios of correlation
functions. We explicitly treat the zero mode contribution and find good control
over the two consecutive limits of $L \to \infty$ and $m_\gamma \to 0$ which are
both guided by effective field theories. We produce numerically compatible
results on two ensembles which differ by a factor of two in the spatial extents,
lending confidence to the chosen approach. We thereby demonstrated that the
absence of power-like finite volume effects in the approach of QED$_{\textrm{M}}$~enables the
use of existing gauge field configurations with $M_\pi L \gtrsim 4$ for
precision predictions of isospin breaking effects and thereby provide a cost
efficient framework.

The results presented here are part of a larger body of work including strong
isospin breaking contributions, the meson and octet spectrum, QED effects on the
$\Omega^-$ baryon mass which is frequently used to set the lattice scale
(e.g.~\cite{Borsanyi:2020mff,Miller:2020evg}), QED corrections to $g_A$,
investigations of the $\Lambda^0-\Sigma^0$ mixing and QED corrections required
for CKM physics. We are also numerically testing the scheme proposed
in~\cite{Bussone:2018ybs} to separate strong and weak isospin breaking
effects. Data has been produced on further ensembles allowing to take a
continuum limit with three lattice spacings at a pion mass of
$310\,\mathrm{MeV}$ and data production for lighter pion masses is ongoing.

\section*{Acknowledgements}
This research used resources of the Oak Ridge Leadership Computing Facility,
which is a DOE Office of Science User Facility supported under Contract
DE-AC05-00OR22725, through the ALCC Program.  This project has received funding
from Marie Sk\l{}odowska-Curie grant 894103 (EU Horizon 2020). MDM and JTT are
partially supported by DFF Research Project 1, grant number 8021-00122B.  The
work of BH and AWL was supported by the LBNL LDRD Program.  AN is supported by
the National Science Foundation CAREER Award Program.  AS is partially supported
by the National Science Foundation grant PHY-1913287.

The QED fields were generated with the code from Ref.~\cite{Endres:2015gda}.
The correlation functions were then computed with Lalibe~\cite{lalibe} on the
\texttt{qedm} branch.  Lalibe links against the Chroma software
suite~\cite{Edwards:2004sx}, which utilizes
QUDA~\cite{Clark:2009wm,Babich:2011np} to perform highly optimized solves of the
MDWF valence propagators on NVIDIA-GPU accelerated compute nodes.

\bibliographystyle{unsrt}
\bibliography{ref}

\end{document}